\documentclass[pra,preprintnumbers,nofootinbib,twocolumn]{revtex4-1}
\usepackage{multirow}
\usepackage[margin=23mm]{geometry}
\usepackage[table,xcdraw]{xcolor}
\usepackage{amsmath}
\usepackage{amsthm}
\usepackage{braket}
\usepackage{hyperref}
\usepackage{graphicx}
\usepackage{qcircuit}
\usepackage[section]{placeins}
\usepackage{enumitem}
\usepackage{comment}
\usepackage{placeins}
\usepackage{nicefrac}
\usepackage{fancyhdr}
\usepackage{soul}

\newcommand{\bqa}{\begin{eqnarray}}
\newcommand{\eqa}{\end{eqnarray}}
    
\newcommand{\be}{\begin{equation}}
\newcommand{\ee}{\end{equation} }

\DeclareMathOperator{\CX}{CX}
\DeclareMathOperator{\R}{R}
\DeclareMathOperator{\U}{U}
\DeclareMathOperator{\VZ}{VZ}

\def\barray{\begin{eqnarray}}
\def\earray{\end{eqnarray}}
\def\beq{\begin{equation}}
\def\eeq{\end{equation}}

\begin{document}

\author{Daniel Bultrini}
\author{Max Hunter Gordon}
\author{Esperanza L\'opez}
\author{Germ\'an Sierra}

\affiliation{Instituto de F\'{\i}sica Te\'{o}rica,  Universidad Aut\'{o}noma de Madrid,  Cantoblanco, 
c/ Nicol\'as Cabrera, 13-15, 28049 Madrid, Spain}

\title{Simple mitigation strategy for a systematic gate error in IBMQ}

\begin{abstract}
    We report the observation and characterisation of a systematic error in the implementation of $U_3$ gates in the IBM quantum computers. The error appears as a consistent shift in one of the angles of the gate, whose magnitude does not correlate with IBM's cited errors calculated using Clifford randomized benchmarking. We propose a simple mitigation procedure, leading to an improvement in the observed value for the CHSH inequality, highlighting the utility of simple mitigation strategies for short-depth quantum circuits.   
\end{abstract} 

\maketitle

\section{Introduction}

    Quantum error correction is essential in the development of fully functional quantum computers. Existing hardware does not meet the requirements to implement fault-tolerant quantum error correction, outside of small preliminary studies \cite{Campagne-Ibarcq2020, Waldherr2014, Kelly2015, Cramer2016}. The accuracy of observables produced by current hardware is therefore limited, but many candidate applications require greater precision to outperform classical methods. For this reason, it is widely regarded that error mitigation will be essential in demonstrating near-term quantum advantage \cite{Preskill2018QuantumBeyond}.
 
    Error mitigation aims to reduce the effect of noise rather than remove it completely. There are many distinct approaches towards this goal, with two common methods being: optimizing quantum circuits through compilation and machine learning \cite{Cincio_2018, Murali, cincio2020machine} and classical post processing. One of the most promising post processing techniques is zero noise extrapolation \cite{Temme_2017} which combines observables evaluated at several controlled noise levels \cite{Dumitrescu_2018, He_2020}, enabling extrapolation to the zero-noise limit. 
    Recently several new mitigation methods have been developed that make use of learning from data sets constructed using quantum circuit data \cite{arm2020learningbased, czarnik2020error} demonstrating the rapid progress in this field. 
 
    Errors occur due to a multitude of factors in both the qubits themselves and the control hardware. Qubits are not completely isolated from their environment, leading to thermal relaxation and the decoherence of their state. Gate errors result from miscalibration or imperfections in the control hardware and their interactions with the qubits. Furthermore, the readout procedure can misidentify or alter the final qubit state such that the measured value does not accurately reflect the collapsed state \cite{Kjaergaard2020}.


    A single qubit pure state can be represented as:
\begin{equation} \label{eq:qubit_state}
    \ket{\psi}=\cos{\frac{\theta}{2}}\ket{0}+e^{i\phi}\sin{\frac{\theta}{2}}\ket{1}.
\end{equation} 
    which can be visualized as a point on the Bloch sphere at polar angle $\theta$ and azimuthal angle $\phi$.

    During computation a given number of one and two qubit gates are performed on a set of qubits. In the zero noise limit this has the effect of changing the state by some unitary operation $\U$. 
    Any unitary is decomposed into the physical gate set of the device, 
    $\mathcal{S}$. When implemented in the IBMQ quantum computers  this set is given by 
    $\mathcal{S}= \{\U_{1}(\omega), \R_{x}(\pm \nicefrac{\pi}{2}), \CX\}$, where $\omega$ is some angle. The gate $\U_{1}(\omega)$ is equivalent to $\R_{z}(\omega)$ up to a global phase factor and is implemented virtually within IBMQ. This is achieved by using frame changes with near perfect execution \cite{McKay2017EfficientComputing} and does not involve the action of any physical quantum gates. A general single qubit unitary 
\begin{equation} \label{eq:u3}
    \U_3(\theta, \phi, \lambda) = \left( \begin{array}{cc} \cos(\theta/2) & -e^{i\lambda}\sin(\theta/2) \\
            e^{i\phi}\sin(\theta/2) & e^{i(\lambda+\phi)}\cos(\theta/2)
     \end{array}\right),
\end{equation}
can be decomposed as follows: 
%
%
\barray \label{eq:mckkaydecomp}
\hspace{-0.6cm}        \U_3(\theta, \phi, \lambda) & = &  \R_{z}(\phi) \R_{x} \!  \left(-\frac{\pi}{2} \right) \! \R_{z}(\theta)\R_{x} \!  \left(\frac{\pi}{2}\right) \!  \R_{z}(\lambda),
\earray 
where the $\R_z$ gates are implemented virtually ($\VZ$),  and the $\R_{x}(\pm \pi/2)$ by a pulse \cite{Qiskit}. 

    Once execution of the required gates is complete, the quantum computer measures the qubits, collapsing the state, and outputs the results. The computation is repeated and a vector of counts $\vec{v}_{exp}$, length $2^{n}$ (where $n$ is the number of qubits), is obtained. Relaxation, imperfect coupling of the readout resonator and signal amplification lead to errors in the measurement process \cite{Kjaergaard2020}. Although major improvements in this area are likely to come from improved hardware, it is possible to mitigate the measurement error through various techniques \cite{Maciejewski_2020}. A simple strategy currently implemented within IBM's Qiskit software \cite{IBM} uses data from calibration circuits to mitigate the error using classical post-processing.  This is achieved using the direct construction of a calibration matrix which for one qubit can be written as:
\begin{equation} \label{eq:cal_mat}
    \boldsymbol{M}_{cal} = 
        \begin{pmatrix}
        p_0&1-p_1 \\
         1-p_0&p_1
        \end{pmatrix},
\end{equation} 
    where $p_0$ and $p_1$ are the probabilities that a prepared $\ket{0}$ is measured as $\ket{0}$ and a prepared state $\ket{1}$ is measured as $\ket{1}$ respectively. This technique can be extended to multi-qubit states using a tensor product or correlated Markov noise approaches \cite{bravyi2020mitigating}. The calibration matrix can also be calculated using maximum likelihood techniques and quantum detector tomography \cite{Chen2019DetectorMeasurement}.

    The calibration matrix can then be used to mitigate errors associated with the readout either directly by (i) inversion or through (ii) bounded minimization. 
\begin{enumerate}[label=(\roman*)]
    \item Inversion is done by inverting the calibration matrix as such: $\boldsymbol{M}^{-1}_{cal}\vec{v}_{exp}=\vec{v}_{th}$,  where $\vec{v}_{exp}, \vec{v}_{th}$ are the experimental and ideal vectors of the counts.
    \item   Bounded minimization uses bounded least squares optimization: $\min_{\vec{v}_{th}}| \boldsymbol{M}_{cal}\vec{v}_{th}-\vec{v}_{exp}|$, where bounds ensure the probabilities calculated from $\vec{v}_{th}$ are positive and correctly normalised. 

\end{enumerate}

    These techniques share the assumption that the error rate in state preparation is much lower than the readout error. This is not without merit as single gate errors cited in IBM, Google and Rigetti are all below 0.5\% while their readout errors are around $1-5\%$ \cite{Kjaergaard2020, Qiskit,RigettiComputing}. Yet, any error in state preparation, especially systematic ones, can lead to an inaccurate calibration matrix. 

    In this paper we highlight a systematic error in the execution of the $\U_{3}$ gate in IBM's cloud based computers, which appears as a shift in the angle $\theta$ when implementing the gate $\U_{3}(\theta, \phi, \lambda)$. 
 We propose to mitigate the previous error using an angular shift in $\theta$ in the $\U_3$ gate
 We illustrate the functionality of this mitigation method by measuring the CHSH inequality on data from a real device.   

\section{Error characterisation}
\subsection{Sweeping a meridian}
F

To explore the reliability of the $\U_3$ gate we applied it to  the $\ket{0}$ state  with  
$\lambda=\nicefrac{\pi}{2}$, $\phi=-\nicefrac{\pi}{2}$ and various angles  $\theta$  in the interval $[0, \pi]$ (see eq. (\ref{eq:u3})). 
    This represents a rotation about the $x$ axis ($\R_{x}(\theta)$) on the Bloch sphere that sweeps a whole meridian.
    The gate is followed by a measurement in the $z$ basis.

\begin{figure}[h!]
\mbox{
    \Qcircuit @C=1em @R=.7em {
        \lstick{\ket{0}} & \gate{\U_3(\theta,-\nicefrac{\pi}{2},\nicefrac{\pi}{2})} & \meter \qw 
    }
    }
    \label{fig:calibration}
\end{figure}

\begin{figure}
    \centering
    \includegraphics[trim={0.2cm 0.5cm 0 0},clip,width=\columnwidth]{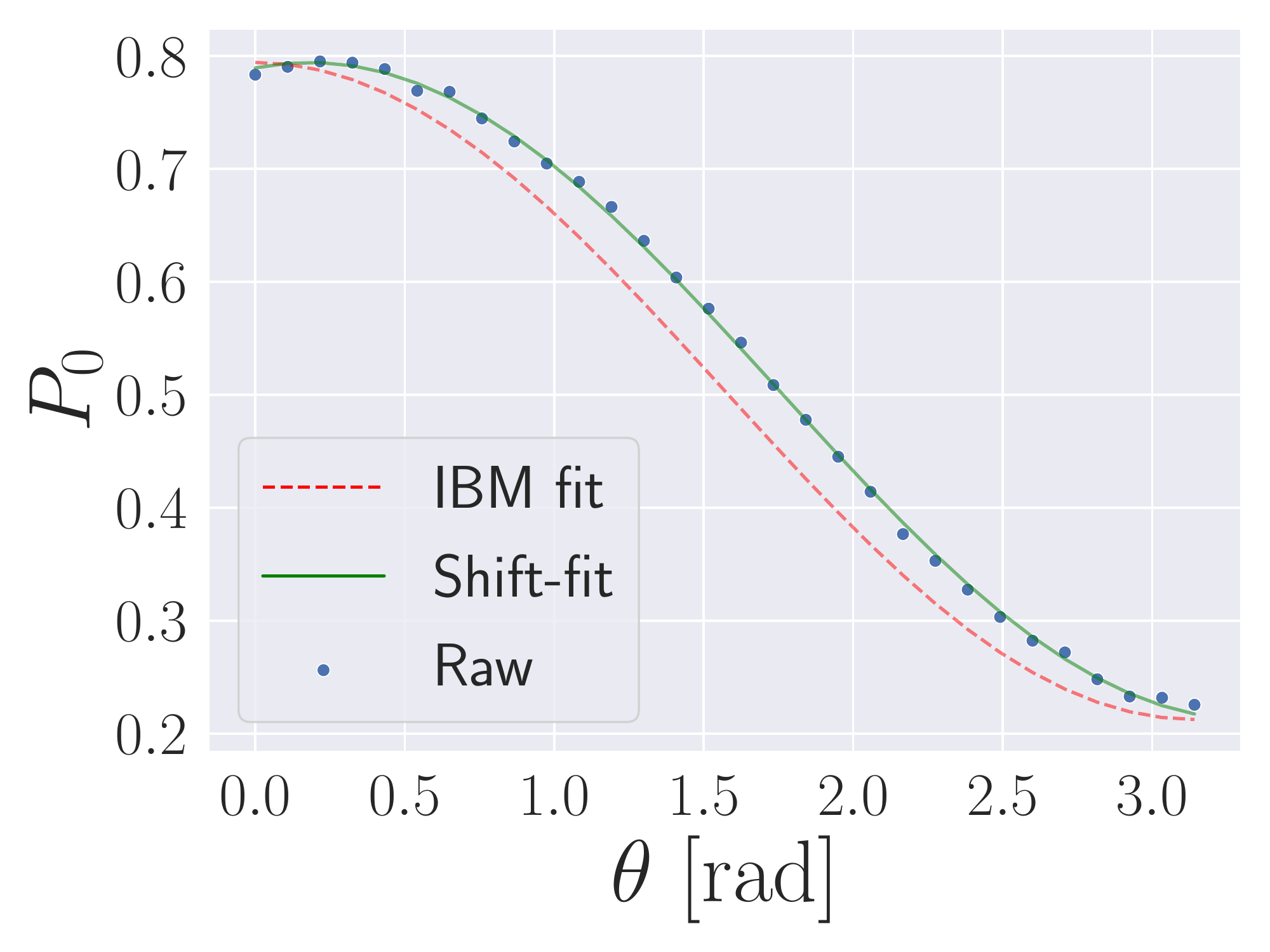}
    \caption{Sweep of $\R_{x}(\theta)$ on Cambridge qubit 9.  The  raw data (blue dots) are fitted with the IBM method (red, dotted) 
    and Shift-fit  with $\alpha=-0.18$ (green, solid). 
    }
    \label{fig:shift-example}
\end{figure}

  IBM's calibration method  consists in measuring the states  $\ket{0}$ and $\ket{1} = R_{x}(\pi)\ket{0}$,   extracting   the values of $p_0$ and $p_1$ to build the matrix $\mathbf{M}_{cal}$ given in \eqref{eq:cal_mat}. The experimental $\ket{0}$ count for any given $\theta$ ($P_{0}(\theta)$), ignoring all errors apart from readout,  
  can be described by 
%
\begin{equation} \label{eq:IBM_fit}
    P_{0}(\theta)=p_0\cos^2{\frac{\theta}{2}}+(1-p_1)\sin^2{\frac{\theta}{2}} \ . 
\end{equation}
We shall  refer to this formula  as the IBM-fit. Observe that \eqref{eq:IBM_fit} reproduces by construction  the experimental
data $p_0$ and $1-p_1$ for $\theta =0$ and $\pi$ respectively.
To test the reliability of this formula we divide $[0, \pi]$ in 30 intervals and measure 
$P_{0}(\theta_i)$ for $\theta_i = \pi i/30$ with $i=0,1, \dots, 30$. 
The results obtained 
for the  qubit  9 of the Cambridge QC, with 8,192 shots per angle, 
 are  plotted in Fig. \ref{fig:shift-example} together with the curve 
\eqref{eq:IBM_fit}. 
One can clearly see  a significant deviation between the experimental  data and the IBM prediction. 
However,  this deviation follows a trend that we   characterize  with the following ansatz
\begin{equation} \label{eq:shift_fit}
    P_{0}(\theta)=p'_0\cos^2{\frac{\theta+\alpha}{2}}+(1-p'_1)\sin^2{\frac{\theta+\alpha}{2}}.
\end{equation}
Here,   the angle  $\theta$ is  shifted by a parameter $\alpha$ that takes small values, as we shall see below. 
The probabilities $p_0$ and $p_1$, appearing in \eqref{eq:IBM_fit},   have  been replaced by $p_0'$ and $p'_1$ 
to allow for a more  accurate  description of the experimental results in the range $\theta \in [0, \pi]$. 
The numerical values of $\alpha, p'_0$ and $p'_1$ are determined using a least-square
fit of the set $\{ P_0(\theta_i) \}_{i=0}^{30}$ using \eqref{eq:shift_fit}. We shall denote this approach as the Shift-fit method. 
Fig. 2 shows that \eqref{eq:shift_fit} provides a much better fit to the data than  \eqref{eq:IBM_fit}.
 
 To quantify the performance of the  fits we use  the coefficient of determination  $R^{2}$ that is defined as 
\begin{equation}
R_{\rm fit}^2 = 1- \frac{  \sum_{n=0}^{30} \left( P^{\rm exp}_0(\theta_n) - P^{\rm fit}_0(\theta_n) \right)^2}{ 
 \sum_{n=0}^{30} \left( P^{\rm exp}_0(\theta_n) - \bar{P}^{\rm exp}_0 \right)^2} 
\end{equation}
where $P^{\rm exp}_0(\theta_n)$ is the experimental probability of the $|0\rangle$  counts at angle $\theta_n$, 
and $\bar{P}^{\rm exp}_0$ its average. The $R^2$ estimator is customarily  expressed in percentages, 
thus a perfect fit, implies a  $R^2_{\rm fit}  \times 100= $ 100\% of predictabilty. 
The data given in Fig. 2 yield an $R^2$ equal  to 97.6\% for the IBM-fit and 99.9\% for the Shift-fit.

\begin{figure}
    \centering
    \includegraphics[width=0.5\textwidth]{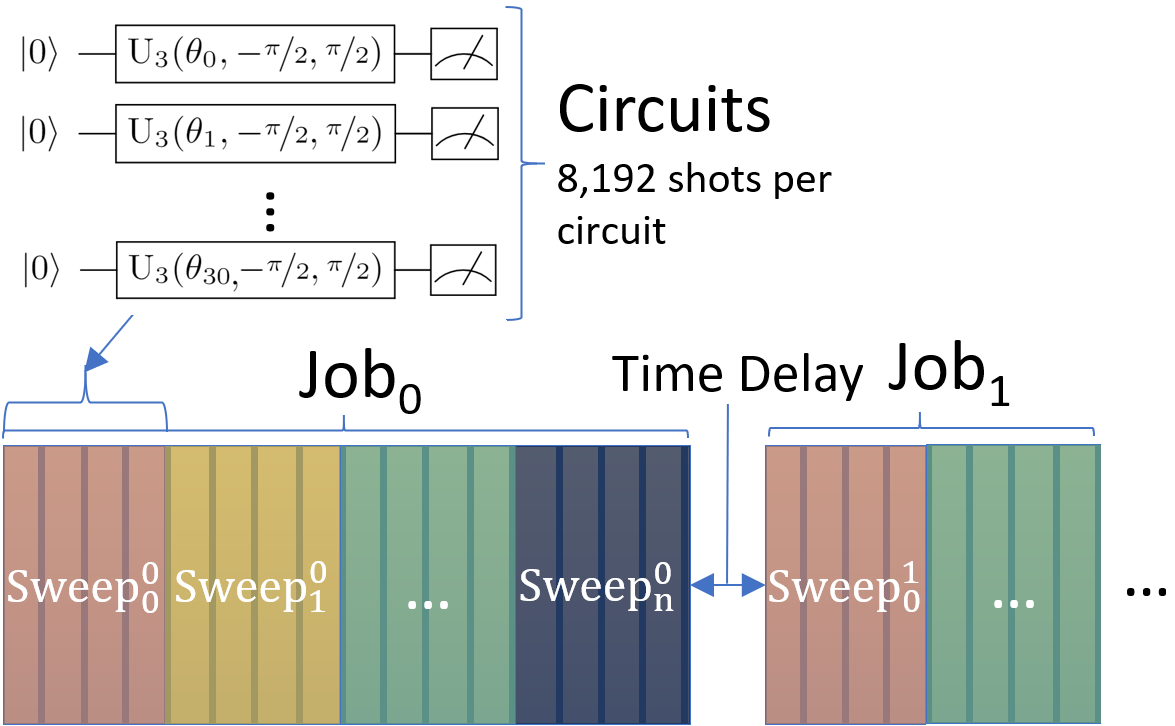}
    \caption{A cartoon describing the way circuits, sweeps and jobs are implemented.
     }
    \label{fig:job_description}
\end{figure}

\subsection{Several sweeps: jobs}

\begin{figure}
    \centering
    \includegraphics[trim={2.8cm 1.2cm 1cm 0.5cm},clip,width=\columnwidth]{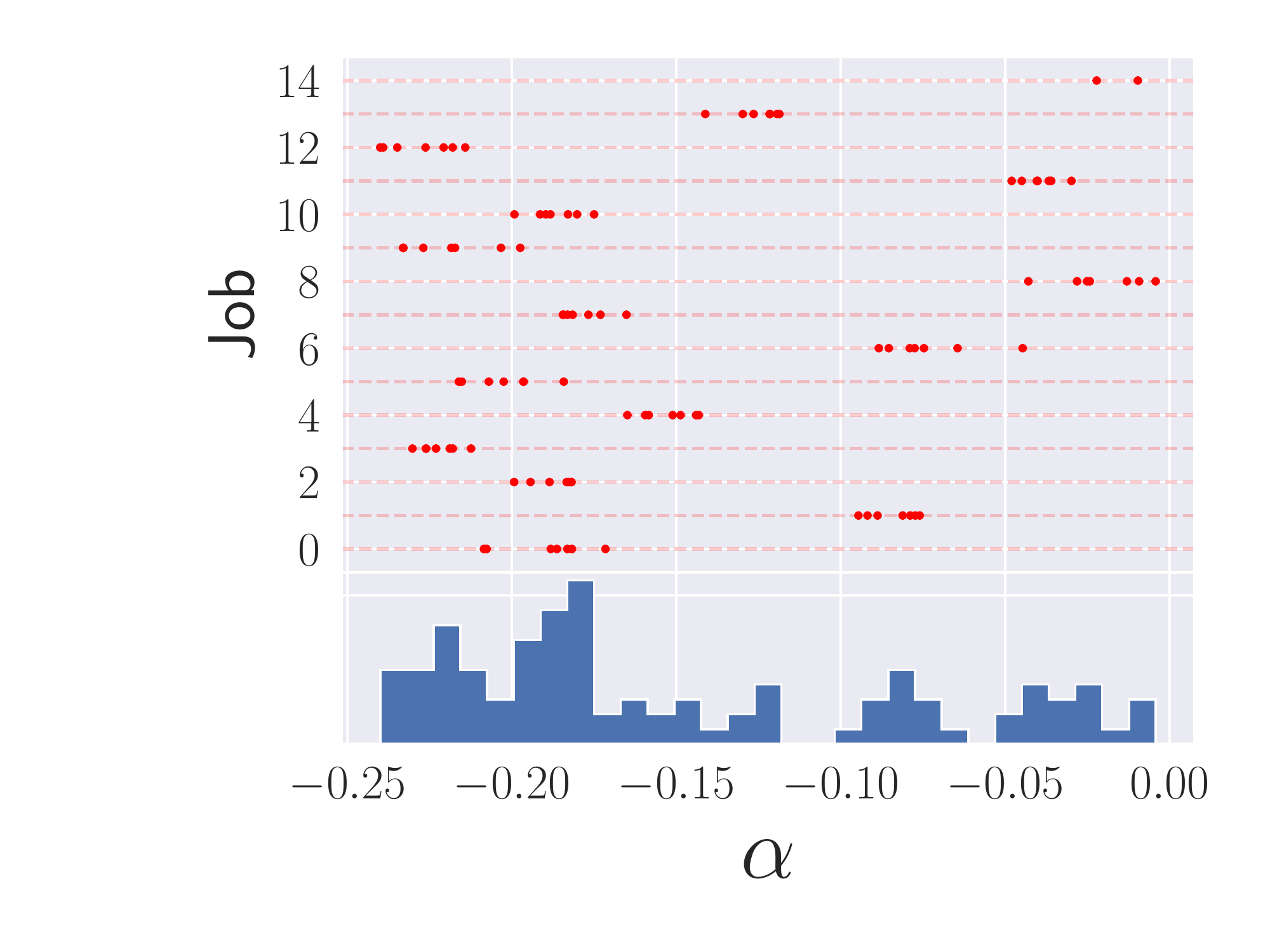}
    \caption{
    Distribution of fitted $\alpha$ values for $100$ $\R_{x}(\theta)$ sweeps for Cambridge qubit $9$. The scatter plot shows the 
    values of $\alpha$ per job over runtime, with the  $15$ different jobs with $n_{s} \in \{2, 5,6, 7\}$
     (100 sweeps)
      denoted by horizontal lines. The bottom displays  a histogram of the  data. 
      }\label{fig:shift-example2}
\end{figure}

The results presented in Fig.  \ref{fig:shift-example}  correspond  to a single sweep of equally spaced
angles  $\theta_n$ along a meridian. 
To assess the reliability of the Shift-fit
method we consider  a set of $n_s$ consecutive sweeps that we denote a job.  
The number of sweeps $n_s$  can depend on the job (see Fig. \ref{fig:job_description}).
A given job is run within a time lapse  where the quantum computer is assumed 
to remain approximately under the same experimental conditions.
The result of  each job is  a set of parameters
$\{ \alpha_s, p'_{0,s}, p'_{1,s} \}_{s=1}^{n_s}$, which  according to the previous assumption,  should be similar.  
Fig. \ref{fig:shift-example2} shows the values of $\alpha$  obtained for 15 jobs,  
amounting  to a total of $100$  sweeps. We notice that:  i) within  each
job the parameter $\alpha$ takes similar values, ii) the average value
of $\alpha$ presents large deviations between jobs,  as shown in the histogram.
Item i) is in  rough agreement with   the stability assumption made above, while item   ii) 
can be attributed to different calibrations during  the time delay  between different jobs. 

The distribution has a mean $\alpha$ of $-0.14(7)$, where the number in brackets is the standard deviation on the last digit shown.  
This mean does not properly reflect how $\alpha$ behaves within a single job, as for example the single run in Fig. \ref{fig:shift-example} whose $\alpha=-0.18$.
We also find that overall the average $R^{2}$ 
for the Shift-fit  and IBM fit are $99.9 \%$ and $97.0 \%$ respectively leading to the conclusion that  including an $\alpha$ shift results in a more accurate description of the raw data in general.
 Finally it is worth noting that 
we have not found correlation between the shift observed and IBM quoted errors.

In table I we collect the results of the observed shift for a selection of qubits in  the devices 
Paris,  Johannesburg,  Rochester,  Cambridge  and  London. 
The chosen qubits are the ones that exhibit the highest  average  values  of $\alpha$. 
The largest twenty average values are provided in the supplementary material. 

We have also explored other meridians with the Shift-fit method and found a negligible dependence on the meridian. Through testing the same qubits in the same job in all the computers with ten equally spaced $\phi$ from $0$ to $2\pi$ we saw a no shifts greater than the standard deviation from the mean and there was no trend of increase with a change in $\phi$.

\begin{table}
\begin{tabular}{ccccc}
Computer                      & Qubit & $\alpha$  & $p'_{0}$ & $p'_{1}$ \\ \hline\hline
Roch                     & 3     & 0.32(6)  & 0.83(4)  & 0.80(2)  \\\hline
\multirow{2}{*}{Johan} & 1     & -0.26(7) & 0.97(1)  & 0.95(2)  \\
                              & 8     & 0.12(5)  & 0.98(2)  & 0.96(2)  \\\hline
Camb                     & 9     & -0.14(7) & 0.82(2)  & 0.81(2)  \\\hline
Lond                      & 2     & -0.12(5) & 0.99(1)  & 0.91(3)  \\\hline
Paris                         & 5     & -0.08(1) & 0.90(1)  & 0.89(2)  \\
\hline
\end{tabular}
\caption{Table showing average parameters from \eqref{eq:shift_fit} fitted to data from 100 sweeps over 10 jobs from different IBM quantum computers. Only qubits with the largest shift $\alpha$ are displayed. 
The standard deviation on the last digit is shown in round brackets after the mean value. }
\label{tab:curve_data}
\end{table}

\subsection{Mitigation}

As explained above,  the parameter $\alpha$ represents a systematic error that affects the
rotation angle $\theta$ of the $\U_3(\theta,\phi,\lambda)$ gate. A naive way to mitigate it 
is to replace $\theta$ by $\theta - \alpha$, expecting that this displacement will compensate the error. 
The corresponding mitigated circuit is
\begin{figure}[h]
\mbox{

    \Qcircuit @C=1em @R=.7em {
        \lstick{\ket{0}} & \gate{\U_3(\theta-\alpha,\phi,\lambda)} & \meter \qw
    }}
\end{figure}

To implement the $\alpha$ mitigation a python software suite was written to perform these calibrations and implement the shift on subsequent experiments \cite{danielmax2020}.
   
\begin{figure}[h]
    \centering
        \includegraphics[width=\columnwidth]{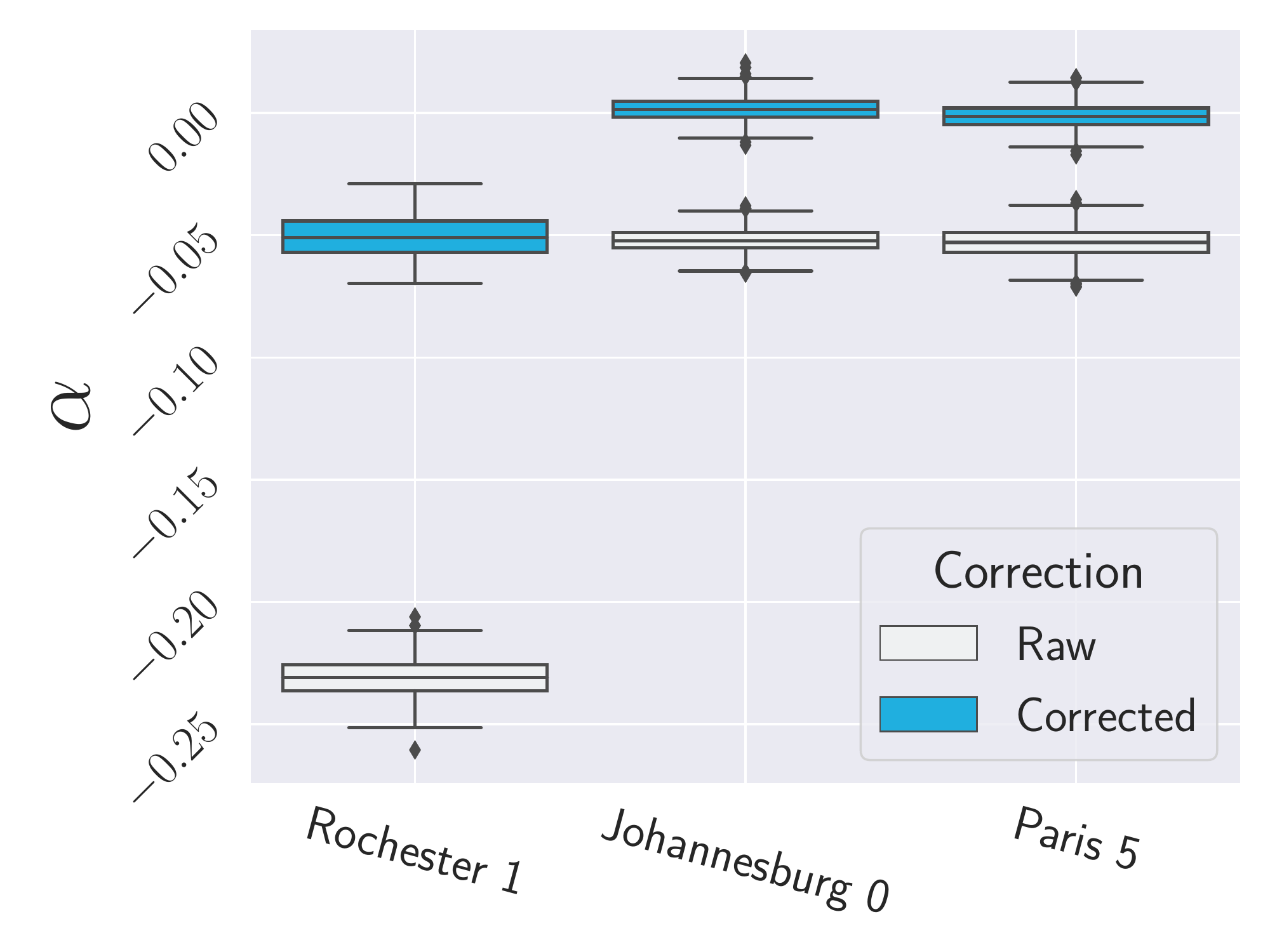}
    \caption{Box plot of the Shift ($\alpha$) determined before (white) and after (blue) mitigation for a subset of qubits from several computers. The box and whiskers encompass 50\% and 95\% of the results respectively, dots represent outliers. 
    Discrepancies between the data for the Paris quantum computer displayed here and that shown in table \ref{tab:curve_data} are due to the results being from different jobs. 
Furthermore, results from some qubits which are displayed in table \ref{tab:curve_data}  are not shown here as they exhibited very small $\alpha$ values at the time of execution, highlighting the large variance of the observed shift between runs.
}\label{fig:correction_applied}
\end{figure}

    Fig. \ref{fig:correction_applied} shows a selection of results. The values of $\alpha$,  
 obtained with this type of mitigated circuit
 are much closer to zero
 that those obtained without the shift.
 The calibration and mitigated rotation were performed with a job with 10 sweeps. 
 The $R^{2}$ values for the Shift-fit were above $99\%$ in all cases. 
 These results assess the effect of the mitigation method.

\subsection{Repeated gates and different initial states}

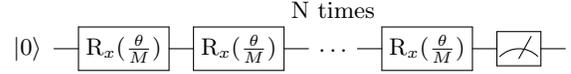
\begin{figure}[h]
    \mbox{
    
        \Qcircuit @C=1em @R=.7em {
          & & & & & \mbox{N times} \\
          &  \lstick{\ket{0}} & \gate{\R_{x}(\frac{\theta}{M})} & \gate{\R_{x}(\frac{\theta}{M})} & \qw& \ldots & & \gate{\R_{x}(\frac{\theta}{M})} & \meter & \qw
        }}
            \caption{Repeated application of rotation gate $\R_{x}(\theta/M)$ to complete at full $\theta$ rotation.}
        \label{fig:spli_application}
    \end{figure}

\begin{figure}
    \centering
    \includegraphics[width=\columnwidth]{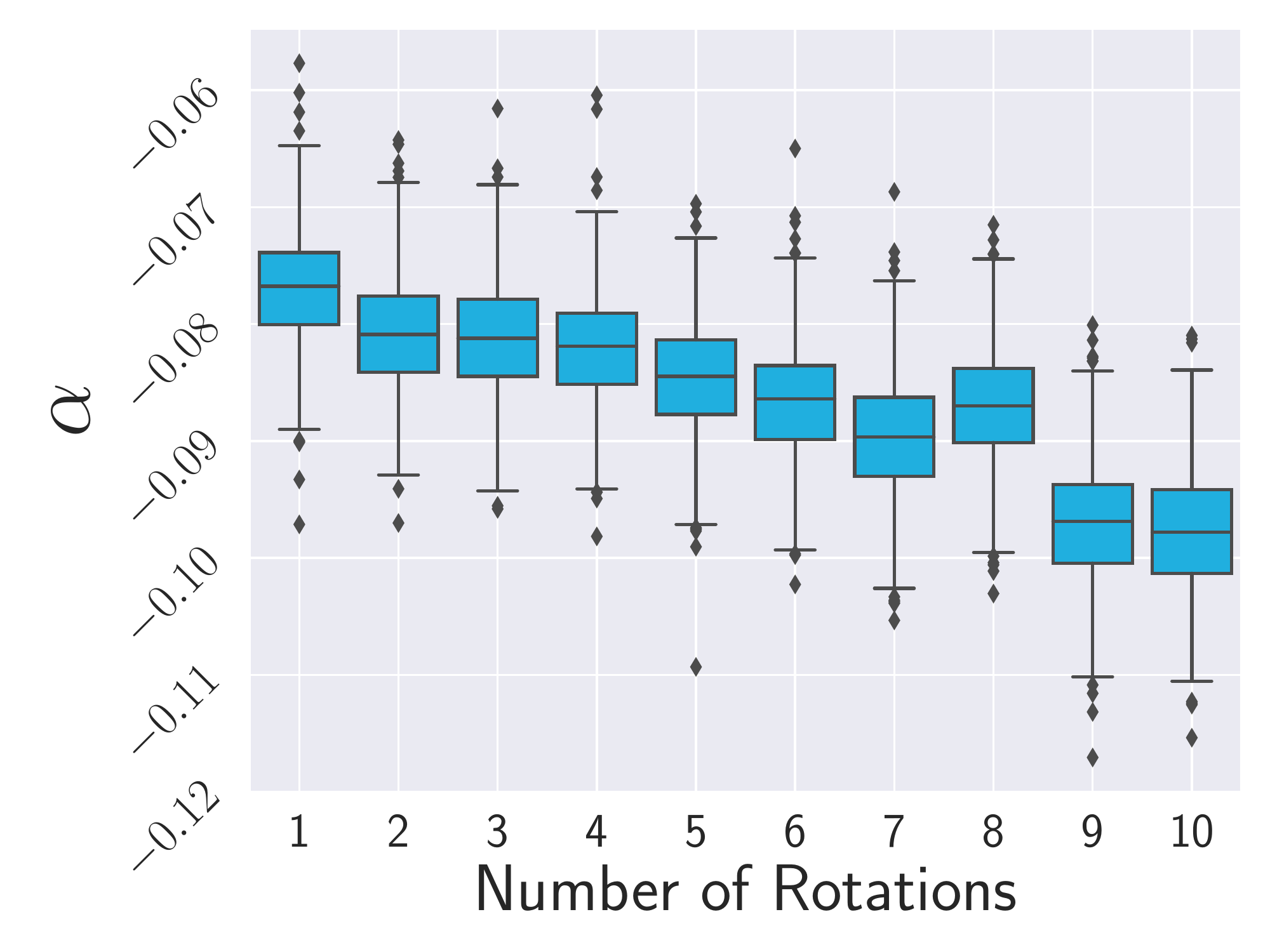}
    \caption{Distribution of $\alpha_M$ using the circuits shown in Fig. \ref{fig:spli_application}. 
We perform 10 sweeps for each value of  $M=1, \dots, 10$, on the 
  Cambridge qubit $9$.  The $R^2$ value does not appreciably decrease when  increasing $M$, 
  implying the fit stays consistent. }\label{fig:repeated_gates}
\end{figure}

We now explore the dependence of the $\alpha$ shift with the number of gates applied in a consecutive sequence.
To this end we decompose a rotation $\R_{x}(\theta)$ into $M$ rotations of angle $\theta/M$, 
as shown in the circuit of Fig. \ref{fig:spli_application}. The results for $M=1, \dots, 10$ are given in  Fig. \ref{fig:repeated_gates}.
We find  that $|\alpha_M|$ increases with $M$, but not linearly as one would naively expect, that is $\alpha_M \simeq M \alpha_1$.  
 All the tested computers returned different trends,  and they changed between jobs 
 even for the same computer. Sometimes a negative $\alpha$  would go closer to zero or further from zero and a positive $\alpha$ would sometimes grow or decrease.
This fact  suggests that the systematic error expressed by  
$\alpha$  has a complex origin that probably involves  several components of the machine.

\begin{figure}
    \centering
    \includegraphics[width=\columnwidth]{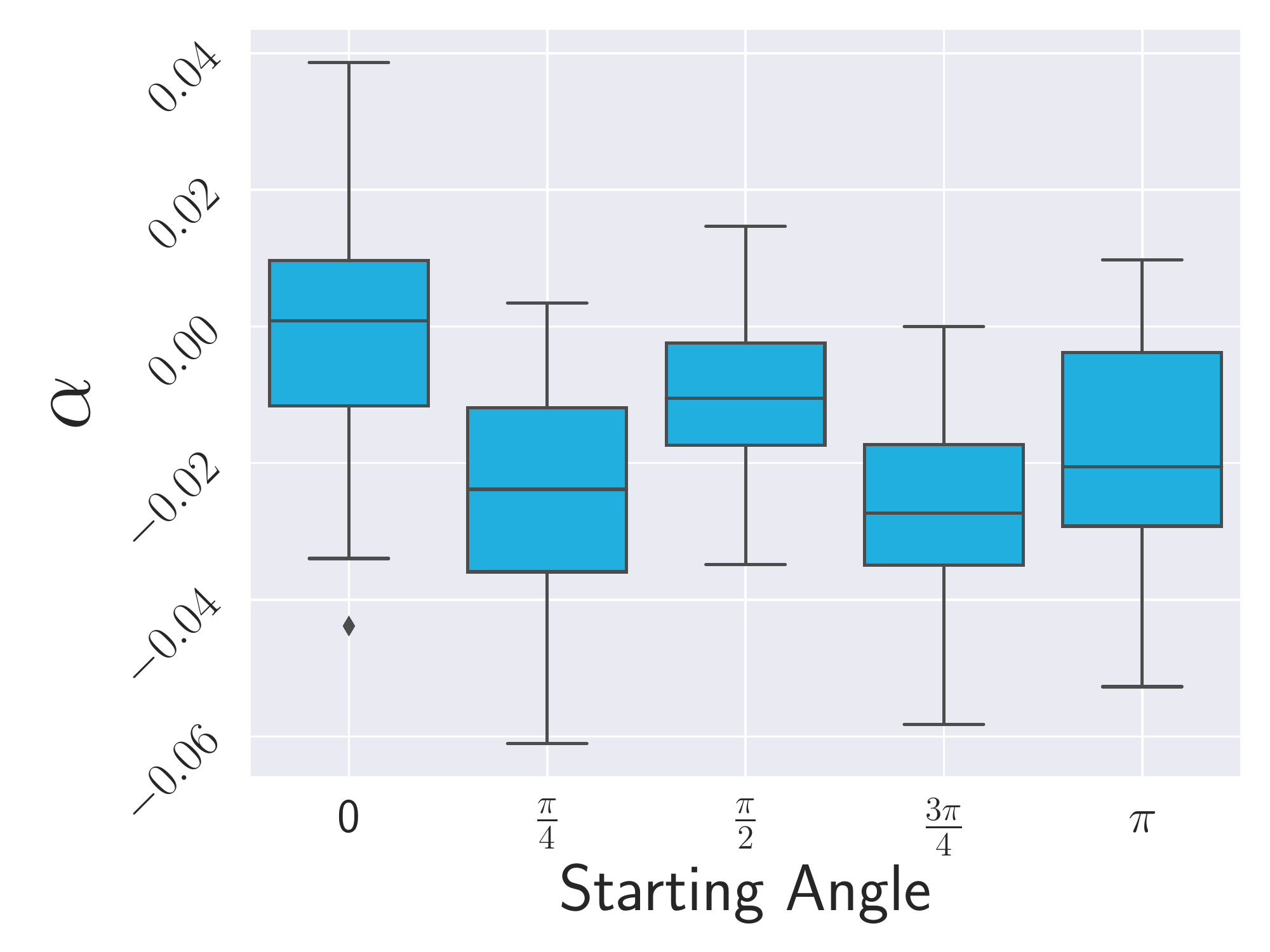}
    \caption{Distribution of $\alpha$ values starting the  sweep from various initial states on Cambridge qubit 9.
    We employ 10 sweeps per state. The observed trend is not fully consistent between computers or qubits, hence the state dependence is not consistent.
 }\label{fig:starting_points}
\end{figure}

We have also studied sweeps starting,  not from $\ket{0}$, 
but from the states obtained acting on $\ket{0}$ with $\R_{x}(\nicefrac{\pi}{4})$, $\R_{x}(\nicefrac{\pi}{2})$ and $\R_{x}(\nicefrac{3\pi}{4})$.
The results plotted  in Fig. \ref{fig:starting_points},  show  a rough agreement of the values of $\alpha$. This suggests the result is not strongly state dependent.

\section{Origin of the Error}

In this section we propose an  explanation of the shift-fit effect based on a potential error
in the implementation  of the gates $\R_{x}(\pm\pi/2)$. In the ideal case these gates are realized as $\exp{( \mp   i t \Omega/2 \ \sigma_X)}$,
where $\Omega$ is the pulse amplitude and $\Omega t = \pi/2$. An off resonance error (ORR) 
in the $\R_{x}$ gate pulse can be modelled as follows  \cite{McKay2017EfficientComputing}:


\barray
&    \R_{x}  (\pm \pi/2,  \delta)  =   {\rm exp} \left[  \frac{ i \pi}{4}  \left(   \mp \sigma_X + \delta \sigma_Z \right) \right]   \\[3mm]         
   &  =     
\left( 
\begin{array}{cc} 
 \cos \frac{\pi d}{4}  - \frac{ i \delta }{d} \sin \frac{\pi d}{4}  &  \mp  \frac{ i}{d} \sin \frac{\pi d}{4}   \\ 
             \mp   \frac{i}{d} \sin  \frac{\pi d}{4}   & \cos \frac{\pi d }{4}  + \frac{ i  \delta}{d}   \sin \frac{\pi d}{4} \\
             \end{array} 
             \right)   &
             \nonumber 
\earray 
where $d = \sqrt{1 + \delta^2}$. Replacing these gates into  \eqref{eq:mckkaydecomp} 
we obtain a gate $U_{3}(\theta, - \frac{\pi}{2}, \frac{\pi}{2}, \delta)$ that includes the  ORR error. 
Finally, we apply  the calibration matrix $\textbf{M}_{cal}$, to obtain the 
 probability  of measuring the $\ket{0}$ state for various angles $\theta$

\begin{equation}
\begin{split}
        P^{\rm ORR}_{0}&(\theta, \delta) = \frac{1 + p_0 - p_1}{2} \\  &  + \frac{p_0 + p_1 -1}{2} \left[ ( 1 - 2 \delta^2) \cos \theta  - 2 \delta \sin \theta \right] \\
        & + O(\delta^3) \ , 
   \end{split}
\end{equation}
where we have assumed that $\delta$ is a small parameter. 
Starting from \ref{eq:shift_fit} and expanding in powers of $\alpha$ gives
\begin{equation}
\begin{split}
        P^{\rm shift}_{0}&(\theta, \delta) = \frac{1 + p_0 - p_1}{2} \\  &  + \frac{p_0 + p_1 -1}{2} \left[ ( 1 - \frac{\alpha^2}{2}) \cos \theta  - \alpha  \sin \theta \right] \\
        & + O(\alpha^3). 
   \end{split}
\end{equation}
These two expression are equivalent up to $O(\delta^{3})$ assuming $\alpha = 2 \delta$ and the using the same calibration matrix. This means that the $\VZ$ gates can indeed be used to correct for this by replacing the $\theta$ parameter in eq. \ref{eq:mckkaydecomp} with $\theta-\alpha$, which is equivalent to altering the $\theta$ in the $\U_3$ gate. 
   
It appears that the shift observed is well described by the appearance of ORR errors in the $R_{x}$ gates. However, upon multiple action of these gates, one would expect the errors to accumulate, resulting in a shift that grows proportionally with the number of applied gates. As previously demonstrated, this is not observed (see Fig. \ref{fig:repeated_gates}). 

We shall show that despite the previous complications, the $\alpha$ mitigation improves observed CHSH inequalities, suggesting the simple mitigation strategy we present could be useful in short-depth circuits.

\section{Evaluating the CHSH inequality}

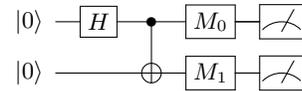
\begin{figure}[h]
\mbox{
    \Qcircuit @C=1em @R=.7em {
      &  \lstick{\ket{0}} & \gate{H} & \ctrl{1} & \gate{M_0} &\meter \qw \\
      &\lstick{\ket{0}}&\qw & \targ & \gate{M_1} &\meter \qw
    }}
\caption{CHSH circuit, $M_{0}$, $M_{1}$ represent the gates required for the basis changes 
to go into the 
    $A_0B_1$, $A_0B'_1$, $A'_0B_1$ and $A'_0B'_1$ bases in order to measure $A=Z$, $A' = X$, $B=\nicefrac{1}{\sqrt{2}}(X+Z)$ and $B'=\nicefrac{1}{\sqrt{2}}(-X+Z)$.}
    \label{fig:CHSH}
\end{figure}

    The CHSH inequality involves running 4 separate circuits which each consist of a Bell state preparation followed by measurements in four appropriately chosen bases (Fig. \ref{fig:CHSH}). It is a quintessential experiment in quantum mechanics demonstrating that quantum correlations cannot be explained classically \cite{Clauser1969ProposedTheories}. The correlation function can be expressed as follows:

\begin{equation}
    C  =  \langle AB \rangle + \langle AB'\rangle +\langle A'B \rangle - \langle A'B' \rangle\\
\end{equation} \label{eq:CHSH}

    where $4$ system observables are shown as $A, A'$ and $B, B'$, these letters simply represent different measurement bases of the bipartite system comprising of $A$ and $B$. $\langle AB\rangle$ is the correlated expectation for two of those observables. For a system with a hidden variable or classical correlations, $|C|$ is bounded at 2. For a system with maximal entanglement, this bound is $2\sqrt{2}$ \cite{Tson1980}.
\begin{table}[]
\begin{tabular}{cccccc}
Computer     & Qubits & $\alpha_0$ & $\alpha_1$ & $C_{raw}$ & $C_{corr}$ \\ \hline\hline
Johan & 6,7    & 0.052(7)   & -0.072(7)  & 2.52(2)   & 2.62(2)    \\
\hline
Lond       & 1,3    & -0.02(1)   & -0.04(1)   & 2.21(2)   & 2.26(2)    \\
Lond       & 1,2    & 0.23(1)    & 0.12(7)    & 2.24(2)   & 2.46(2)    \\
\hline
Roch   & 3,4    & -0.07(1)   & -0.012(7)  & 2.22(4)   & 2.26(5)    \\
Roch    & 44,51  & 0.032(5)   & -0.026(4)  & 1.5(1)    & 1.82(8)    \\
Roch    & 48,52  & 0.006(3)   & -0.05(1)   & 1.23(3)   & 1.25(3)    \\
\hline
Paris        & 8,9    & 0.007(7)   & 0.012(2)   & 2.51(2)   & 2.51(2)   \\
\hline
Camb & 9,10 & -0.02(1) & 0.011(5) & 2.06(1) & 2.085(9) \\
\hline
\end{tabular}
\caption{Shift values and correlation functions showing raw and $\alpha$ mitigated implementations of the CHSH inequality circuits for 819,200 shots per basis. Qubits with local connectivity were chosen to minimize the depth of the circuits necessary. The  calibration of $\alpha$ was calculated with $10$ repetitions. In all cases where there is a significant shift we see either a statistical improvement in the measured value for $C$.}\label{tab:CHSHIN}
\end{table}

    In general the measured mitigated correlations are closer to the theoretical limit as  in table \ref{tab:CHSHIN}, with the least improved cases appearing when $\alpha$ is very small in one or both qubits. Therefore, using a simple mitigation strategy can improve measured quantities in a real device.

    How this improvement scales with depth and number of qubits in the circuit is an important consideration. We have shown the shift effect does not appear to be consistent with increasing depth as seen in \ref{fig:repeated_gates}. However, when increasing the system size a set of calibration circuits could be run on each qubit to determine the $\alpha$ shift whose effect could then be mitigated as outlined above.

\section{Discussion and Conclusion}
    In this paper we have highlighted the existence of a systematic error, which appears as an angular shift ($\alpha$) in the parameter  $\theta$ of the $\U_3$ gate, and demonstrated its effects can be mitigated by performing a simple 
    calibration before running a set of jobs. This shift was shown to bare characteristics of an ORR error. Therefore, it is now possible to mitigate this component of the total error irrespective of the readout error and other errors. This leads to an increased performance on our benchmark circuits to calculate the CHSH inequality. We found that the systematic shifts are consistent over the time span of a few successive jobs, but not over larger stretches of time. 

    As the ORR error can be corrected through the use of $\VZ$ gates, the change in the $\theta$ parameter of the $\U_3$ gate does just this \cite{McKay2017EfficientComputing}. Although using the 'open pulse' capabilities of some IBMQ quantum computers and finely tuning the $R_{x}$ pulses would result in similar improvements, this is a more complicated procedure and may not completely remove the ORR effect.   
    
    We have also shown that although these errors can be corrected for single gates, the application of multiple gates to a single qubit does not follow the expected relation 
    from the ORR treatment which imply a linear growth in the shift with multiple gates. The origin of this behaviour remains an open question 
     and further investigation is left to future work. Despite this, applying this correction still yielded improved results in the CHSH inequalities. 

    Any simple mitigation strategy can only improve the fidelity of calculations by a small factor. Yet, a modest increase in fidelity for a small upfront computation may be worth the extra time. Although this method could not be applied to deep circuits we envision it could be useful for many qubit, short-depth quantum circuits, especially if combined with other mitigation techniques.

\section*{Acknowledgements}

We would like to thank Diego  Garc\'{\i}a-Mart\'{\i}n and Pol Forn  for conversations.
We also thank the IBM Quantum team for making multiple devices available 
via the IBM Quantum Experience. 
The access to the IBM Quantum Experience has been provided by the CSIC IBM Q Hub. 
We acknowledge support from La Caixa Foundation (DB, MH), 
European Union's Horizon 2020 research and innovation programme under the Marie Skodowska-Curie grant agreement No. 71367 
(DB).  This work has also been financed by the Spanish 
grants PGC2018-095862-B-C21, QUITEMAD+ S2013/ICE-2801, SEV-2016-0597 of the
"Centro de Excelencia Severo Ochoa"  Programme and the CSIC Research Platform on
Quantum Technologies PTI-001.

\bibliography{bibliography}

\onecolumngrid
\newpage

\section*{Supplementary Material}

\subsection{Coefficient of determination, $R^2$}\label{sect:determination}
The coefficient of determination, $R^2$, is defined as 
\begin{equation}
    R^2 = 1 - {\frac{SS_{\rm res}}{SS_{\rm tot}}}
\end{equation}
Where the total sum of squares $SS_\text{tot}$ and total sum of residuals $SS_\text{res}$ are 
\begin{align}
    & SS_\text{tot}=\sum_i (y_i-\bar{y})^2\\
    & SS_\text{res}=\sum_i (y_i - f_i)^2,
\end{align}
with $y_i$ being a particular data point, $f_i$ being the prediction of $y_i$ and $\bar{y}$ the average of the observed data. If $R^2=1$, the fit is an exact match to the experimental data while anything lower implies a progressively worse fit. 

In total the statistics of the goodness of fit of our proposed shift with respect to IBM and the ideal curve (setting $p_0=p_1=1$ and $\alpha=0$) are tabulated below for an aggregate of all of the sweeps over all computers. 

\begin{table}[h]
\begin{tabular}{l|lll}
\textbf{$R^2$ values for:} & \textbf{Ideal} & \textbf{Shift-fit} & \textbf{IBM} \\ \hline
Mean                       & 0.576          & 0.9995             & 0.9737       \\
STD                        & 0.153          & 0.0002             & 0.0175       \\
Min                        & 0.310          & 0.9989             & 0.9434       \\
Max                        & 0.794          & 0.9998             & 0.9996      
\end{tabular}
\end{table}

Furthermore, we ascertained that there was no correlation between the alpha values and the cited IBM error rate by ordering the size of the errors for a given computer's qubits by magnitude and comparing them to the magnitude of $\alpha$ associated with a given job. There was no polynomial (up to order 4) which gave any appreciable $R^2$ value for any computer.
\newpage
\subsection{Largest observed shift values}
    The table below shows the fitted data for $20$ qubits with the largest average $\alpha$ after $100$ sweeps, with exception of Rochester at $10$ sweeps due to the large number of qubits. This process was carried out on the Cambridge, London, Rochester, Paris and Johannesburg computers. 
\begin{table}[h]
\begin{tabular}{ccccc}
Computer & Qubit & $\alpha$ & $p_0$ & $p_1$ \\ \hline \hline
Roch     & 3      & 0.32(6)    & 0.83(4)    & 0.80(3)   \\ \hline
Roch     & 51     & -0.29(3)   & 0.80(1)    & 0.72(4)   \\ \hline
Johan    & 1      & -0.26(7)   & 0.97(1)    & 0.95(2)   \\ \hline
Roch     & 52     & -0.19(2)   & 0.88(1)    & 0.81(4)   \\ \hline
Roch     & 30     & 0.16(2)    & 0.85(2)    & 0.87(3)   \\ \hline
Camb     & 9      & -0.14(7)   & 0.82(2)    & 0.81(2)   \\ \hline
Roch     & 35     & -0.13(2)   & 0.85(1)    & 0.85(3)   \\ \hline
Roch     & 8      & 0.13(2)    & 0.91(1)    & 0.89(3)   \\ \hline
Roch     & 12     & 0.13(1)    & 0.89(1)    & 0.89(3)   \\ \hline
Roch     & 13     & 0.12(6)    & 0.66(2)    & 0.65(5)   \\ \hline
Lond     & 2      & -0.12(5)   & 0.99(1)    & 0.91(3)   \\ \hline
Roch     & 2      & -0.11(2)   & 0.88(1)    & 0.86(3)   \\ \hline
Johan    & 8      & 0.11(1)    & 0.98(1)    & 0.96(2)   \\ \hline
Johan    & 10     & 0.11(1)    & 0.96(1)    & 0.94(2)   \\ \hline
Johan    & 9      & -0.10(1)   & 0.96(1)    & 0.94(2)   \\ \hline
Roch     & 41     & 0.10(1)    & 0.97(1)    & 0.93(2)   \\ \hline
Johan    & 3      & -0.10(1)   & 0.96(1)    & 0.96(2)   \\ \hline
Roch     & 23     & 0.10(2)    & 0.87(2)    & 0.84(3)   \\ \hline
Johan    & 0      & 0.09(1)    & 0.94(1)    & 0.92(2)   \\ \hline
Roch     & 27     & -0.09(2)   & 0.89(1)    & 0.92(3)   \\ \hline
Johan    & 7      & -0.09(1)   & 0.98(1)    & 0.96(2)   \\ \hline
\end{tabular}
\caption{Largest $20$ shift values found in the computers that were investigated. The parameters correspond to those shown in \ref{eq:shift_fit}. This was repeated $10$ times and errors show the standard deviation, with the error in the last digit shown in brackets.}
\end{table}

\end{document}